\documentclass[aps,pre,twocolumn,groupedaddress, amsmath, amssymb]{revtex4-1}
 
\newcommand{\erf}{\ensuremath{\textrm{erf}}}

\newcommand{\cc}{\textrm{col}}

\newcommand{\be}{\begin{eqnarray}}
\newcommand{\ee}{\end{eqnarray}}
\newcommand{\ben}{\begin{eqnarray*}}
\newcommand{\een}{\end{eqnarray*}}

\newcommand{\vc}{\mathbf}
\usepackage{graphicx}
\usepackage{color}

\begin{document}


\title{Transport Regimes Spanning Magnetization-Coupling Phase Space}


\author{Scott D.\ Baalrud$^{1}$}
\author{J\'{e}r\^{o}me Daligault$^2$}
\affiliation{$^1$Department of Physics and Astronomy, University of Iowa, Iowa City, Iowa 52242, USA}
\affiliation{$^2$Theoretical Division, Los Alamos National Laboratory, Los Alamos, New Mexico 87545, USA}


\date{\today}

\begin{abstract}
The manner in which transport properties vary over the entire parameter-space of coupling and magnetization strength is explored for the first time. Four regimes are identified based on the relative size of the gyroradius compared to other fundamental length scales: the collision mean free path, Debye length, distance of closest approach and interparticle spacing. Molecular dynamics simulations of self-diffusion and temperature anisotropy relaxation spanning the parameter space are found to agree well with the predicted boundaries. Comparison with existing theories reveals regimes where they succeed, where they fail, and where no theory has yet been developed. 
\end{abstract}




\maketitle


\section{Introduction}

One of the most fundamental characteristics of plasmas is the ability for magnetic fields to strongly influence the transport of particles, momentum and energy. Using magnetic fields to control these properties is the basis for many applications, ranging from magnetic fusion energy to charged particle traps. In addition to magnetization, transport properties are fundamentally influenced by  the strength of Coulomb interactions.
Although great progress has been achieved in understanding collisional transport processes in asymptotic regimes of either weak or strong Coulomb coupling, or weak or strong magnetization, how these regimes merge, and what determines their boundaries, remains an open question. A better understanding of the conditions where reliable theory does or does not exist is critical to progress in several modern research areas that encounter plasmas spanning vast ranges of both coupling and magnetization strengths, including inertial confinement fusion~\cite{gome:14}, white dwarf and neutron stars~\cite{valy:14,Potekhin2015}, charged particle traps \cite{Danielson2015}, and fundamental physics experiments such as dusty~\cite{thom:12,boni:13}, ultracold~\cite{zhan:08}, and nonneutral plasmas~\cite{beck:96}. 

Here, each of the transport regimes that exist in the phase space of magnetization and coupling strength are identified. 
Molecular dynamics (MD) simulations of self-diffusion and temperature anisotropy relaxation are used to quantify how properties smoothly transition from one regime to another, sometimes over decades, in this parameter space. 
Here, ``transport regime'' refers to the conditions of temperature, density and magnetic field strength at which transport properties
arise from identifiable underlying physical mechanisms.
To quantify these regimes, we focus on the one-component plasma (OCP) model~\cite{baus:80,ott:11}. 
Experimental and naturally occurring plasmas are often complicated systems in which the magnetization and coupling strength of 
electrons and ions can differ widely, requiring multi-component models. 
MD simulations of the OCP allow for first-principles computations that are not possible with more complete models, enabling rigorous tests of transport theory. 
The magnetized OCP is characterized by just two dimensionless parameters: the coupling strength $\Gamma\!=\!(e^2/a)/(k_BT)$, which is the ratio of the Coulomb potential energy at the average interparticle spacing $a\!=\! (3/4\pi n)^{1/3}$ to the average kinetic energy, and the magnetic field strength $\beta = \omega_c/\omega_p$, which is the ratio of the gyrofrequency $\omega_c = eB/mc$ to the plasma frequency $\omega_p = \sqrt{4\pi e^2n/m}$. 

\begin{figure}
\includegraphics[width=8.0cm]{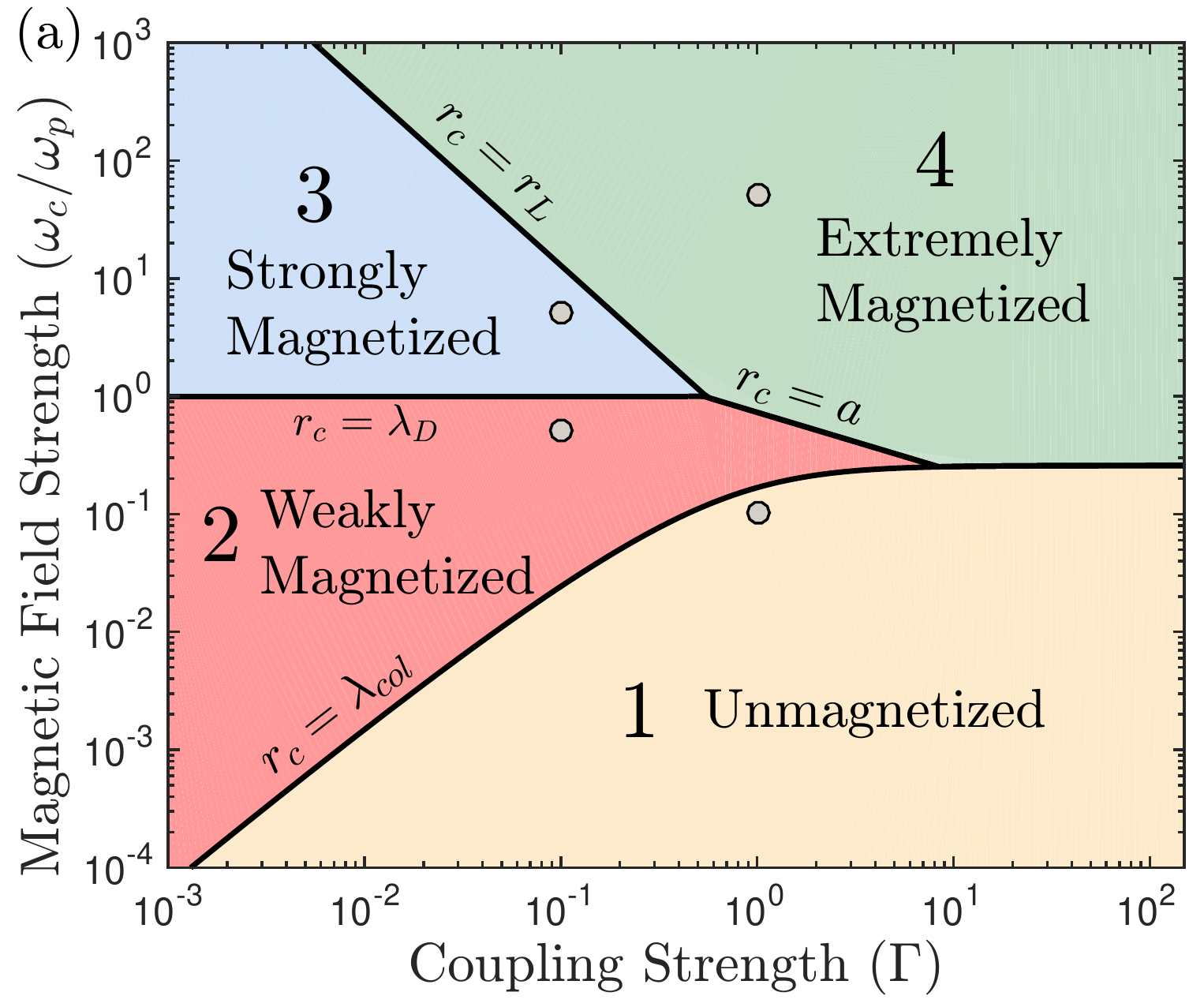}
\includegraphics[width=8.0cm]{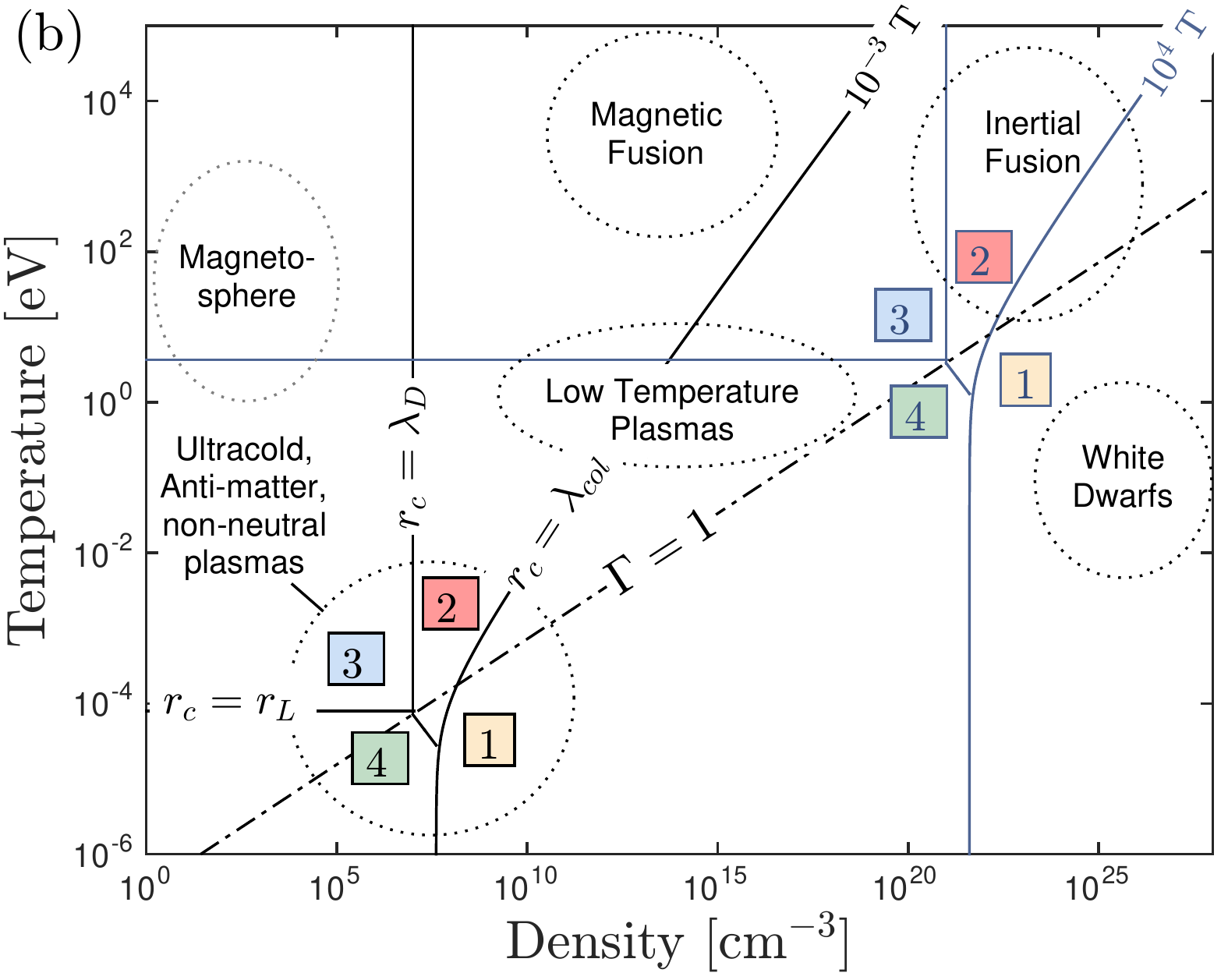}
\caption{(a) Magnetization-coupling phase space indicating the regimes described in Sec.~\ref{sec:regimes}, and the parameter location for each of the videos provided as supplementary material: $(\Gamma,\beta) = (1,0.1)$, (0.1,0.5), (0.1,5) and (1,50). (b) A projection of the four regimes from panel (a) onto the temperature-density plane for two magnetic field strengths: $10^{-3}$ T (black lines) and $10^4$ T (light blue lines). Here, the electron mass was used. }
\label{fg:regimes_sketch}
\end{figure}

The influence of magnetization on collisional transport in weakly coupled plasmas ($\Gamma\ll 1$) has been the subject of numerous works throughout the years \cite{rost:60b,sili:63,brag:65,ichi:70,daws:71,okud:72,mont:74,krom:76,marc:84,cohe:84,sutt:87,onei:83,onei:85,ware:89,glin:92,Dubin1998,ners:07,dubi:14}.
Four regimes have been identified depending on the magnitude of the gyroradius $r_c =\sqrt{k_BT/m}/\omega_c$ compared to, with increasing $\beta$, 
the Coulomb collision mean free path $\lambda_\cc$, the Debye length $\lambda_D = \sqrt{k_BT/(4\pi e^2 n)}$ and the distance of closest approach (i.e., the Landau length) times $\sqrt{2}$: $r_L = \sqrt{2} e^2/k_BT$~\cite{rl:footnote}. 
The boundaries of these regimes are shown in Fig.~\ref{fg:regimes_sketch}a in the $\Gamma-\beta$ plane (see black lines at $\Gamma\ll 1$).
By extrapolating these lines towards larger $\Gamma$ values, the figure immediately suggests a remarkable result: the four regimes relevant to weakly coupled plasmas collapse with increasing $\Gamma$ into two regimes.
By extending the simple arguments based on the ratio of length scales characterizing microphysical processes, we hypothesize that the boundaries identified in Fig.~\ref{fg:regimes_sketch} define four transport regimes in the $\Gamma-\beta$ plane. 
One of these is the unmagnetized regime, where transport rates are expected to be independent of magnetic field strength. 
The other three are magnetized plasma regimes, and are labeled according to the increasing influence of the magnetic field (weak, strong and extreme magnetization). 

Figure~\ref{fg:regimes_sketch}b shows the same regimes in terms of density and temperature at two magnetic field strengths, showing that the boundary between weak and strong magnetization depends only on the density, while the boundary between strong and extreme magnetization depends only on the temperature. 
The plots were obtained using the electron mass. 
Equivalent figures for ions are obtained by multiplying the magnetic field values by the square root of the mass ratio: $\sqrt{m_i/m_e}$.

This paper is organized as follows. 
Section~\ref{sec:regimes} describes the predicted regime boundaries based on the length scale arguments given above. 
Section~\ref{sec:MD} describes the MD simulation method used to determine the self-diffusion coefficients and temperature anisotropy relaxation rate of the OCP. 
Results showing agreement between the MD simulations and predicted regime boundaries are provided in Sec.~\ref{sec:results}, along with a comparison with theoretical predictions in regimes where predictions are available. 

\section{Regimes~\label{sec:regimes}} 

\subsection{Unmagnetized $(r_c\!\gtrsim\!\lambda_\cc)$} 

When the gyroradius exceeds the Coulomb collision mean free path, the magnetic field does not significantly influence collisional transport. 
To estimate $\lambda_\cc$ beyond the weakly coupled regime, we apply the effective potential theory (EPT)~\cite{baal:13,baal:15}, which was recently shown to accurately extend traditional plasma transport theory into the strong coupling regime. 
For the OCP, the EPT predicts $\lambda_\cc=\lambda_D/0.32 \Gamma^{3/2} \Xi(\Gamma)$, where $\Xi(\Gamma)$ is a generalized Coulomb logarithm dependent only on $\Gamma$, 
so that regime 1 is defined by
\begin{equation}
\beta \lesssim 0.32 \Gamma^{3/2} \Xi(\Gamma) \,. \label{eq:R1}
\end{equation}
The calculation of $\Xi$ is explained in \cite{baal:13,baal:15}, and in Appendix \ref{appendix}. An expression accurate over the range of coupling strength considered here is $\Xi(\Gamma)= 0.65 \ln [1+2.15/(\sqrt{3}\Gamma^{3/2})]$. This model gives predictions for both self-diffusion coefficients and temperature relaxation rates of the OCP that are in excellent agreement with MD simulations.

\subsection{Weakly Magnetized ($\max\lbrace \lambda_D, \bar{a} \rbrace\!\lesssim\! r_c \!\lesssim\! \lambda_\cc$)}

In this regime, commonly called the Braginskii regime~\cite{brag:65}, the plasma is magnetized but the gyroradius is greater than the distance over which Coulomb interactions occur: the maximum of the Debye length, $\lambda_D$, or an approximate interparticle spacing, $\bar{a}=a/\sqrt{2}$~\cite{a:footnote}. 
Here, the magnetic field can influence the macroscopic transport coefficients, because the distribution function is more easily distorted along the magnetic field than across it, but not the microphysical process of scattering. 
Thus, in a kinetic theory, magnetization is expected to influence the convective terms, but not the collision operator \cite{brag:65}.  
For the OCP,  regime 2 is defined by
\begin{equation}
0.32 \Gamma^{3/2} \Xi (\Gamma) \lesssim \beta \lesssim \min \lbrace 1, \sqrt{2/(3\Gamma)}  \rbrace  . \label{eq:R2}
\end{equation}
This is the most common regime in laboratory and magnetic fusion experiments; see Fig.~\ref{fg:regimes_sketch}b. 
It is often referred to simply as the magnetized plasma regime, but here we label it ``weakly magnetized'' to distinguish it from the two distinct regimes described below. 

\subsection{Strongly Magnetized ($r_L \!\lesssim\! r_c \!\lesssim\! \lambda_D$)}

In this regime, the plasma is so strongly magnetized that the gyromotion occurs on the same microphysical length scale as scattering, yet the gyroradius remains larger than the Landau length, $r_L$. 
This regime is only relevant to weakly coupled plasmas because $r_L$ and $\lambda_D$ merge at $\Gamma \simeq 0.5$. 
For the OCP, regime 3 is defined by
\begin{equation}
1 \lesssim \beta \lesssim 1/\sqrt{6\Gamma^3} . \label{eq:R3}
\end{equation}
Since gyromotion occurs at the microscale in this regime, one expects that magnetization influences both the convective and collision terms in the kinetic theory~\cite{onei:83}. 

\subsection{Extremely Magnetized ($r_c \!\lesssim\! \min \lbrace r_L, \bar{a}, \lambda_\cc \rbrace$)} 

When the magnetic field is so strong that the gyroradius becomes the smallest length scale in the system, the magnetic field has an extreme influence on transport. 
This occurs if the gyroradius is shorter than: $r_L$ at weak coupling, $\bar{a}$ at moderate coupling and $\lambda_\cc$ at strong coupling. 
For the OCP, regime 4 is defined by
\begin{equation}
\max \lbrace 1/\sqrt{6\Gamma^3} , \sqrt{2/(3\Gamma)}, 0.32 \Gamma^{3/2} \Xi \rbrace\lesssim \beta. \label{eq:R4}
\end{equation}
In this regime, the gyroradius is so small that the charged particle motion is nearly one-dimensional. 
Videos show particle motion along the magnetic field until encountering a neighbor on a nearly field line, followed by a 180$^\circ$ scattering event \cite{sup_footnote}.  

\section{Molecular Dynamics Simulations~\label{sec:MD}} 

MD simulations of self-diffusion parallel and perpendicular to the magnetic field, as well as the rate of relaxation of a temperature anisotropy were used to test the suggested regime boundaries. 
These processes were chosen because they represent momentum and energy exchange processes, respectively, which show quite different dependencies on the regime boundaries. 

The simulations were performed as follows.
$N$ charged particles interacting through the pure Coulomb interaction in a uniform, neutralizing background were placed in a cubic box of volume $V$.
Periodic conditions were imposed on all boundaries.
The particle trajectories were determined by solving Newton's equations of motion with the time integrator of Spreiter and Walter \cite{SpreiterWalter1999} that extends the traditional Verlet integrator \cite{FrenkelSmit} to handle arbitrarily strong static homogeneous external magnetic fields.
The force on an ion that results from its interaction with the external magnetic field, with the ions in the simulation box and with those in the periodically replicated cells were calculated using the Ewald summation technique.
For numerical efficiency, the Ewald sum was calculated with a parallel implementation of the particle-particle-particle-mesh ($\rm P^3M$) method that simultaneously provides high resolution for individual encounters combined with rapid, mesh-based, long range force calculations \cite{HockneyEastwood}.
Time was normalized to the plasma frequency $\omega_p$.
The integration time step $\delta t$ and the number of particles $N$ given below were chosen in order to conserve energy to $<10^{-5}$, and to ensure high enough collision ability in the simulation cell (the calculations are more demanding at small coupling due to long collision mean-free path).
Specifically, for $\Gamma\leq 0.1$, we used $N=10^5$ and $\delta t=10^{-3}/\omega_p$; for $0.1<\Gamma<1$, $N=50000$ and $\delta t=0.01/\omega_p$; for $\Gamma\geq 1$, $N=5000$ and $\delta=0.01/\omega_p$.
 
Each simulation consisted of an equilibration phase of length $t_{\textrm{eq}}=N_{\textrm{eq}}\delta t =1000/\omega_p$ followed by a main run phase of length $t_{\textrm{run}}$.
The initial particle positions at time $t=-t_{\textrm{eq}}$ were assigned randomly in the simulation box, with a small region surrounding each particle excluded to avoid initial explosion.
The initial particle velocities ${\bf v}_i$ were sampled from a Maxwell-Boltzmann distribution at the desired temperature $T$, i.e. at the desired value of $\Gamma$.
During the equilibration phase, velocity scaling \cite{FrenkelSmit} was used to maintain the desired $\Gamma$ value.
The main run phase varied depending on the physical property under study.


\subsection{Self-Diffusion Coefficients}

Following the equilibration phase, velocity scaling was turned off and the external magnetic field ${\bf B}=B\hat{z}$ was turned on. 
The system was then left to evolve freely in the microcanonical ensemble for a duration $t_{\textrm{run}}=5242.88/\omega_p $. 
During this phase, the particle positions and velocities were recorded at every timestep.
After the simulation was completed, the parallel and perpendicular self-diffusion coefficients were evaluated from the Green- Kubo relation,
\begin{eqnarray}
D_\parallel&=&\frac{1}{N}\sum_{i=1}^N\int_0^\infty{\langle v_{z,i}(t)v_{z,i}(0)\rangle dt} \label{Dparallel}\\
D_\perp&=&\frac{1}{2N}\sum_{i=1}^N\int_0^\infty{\left[\langle v_{x,i}(t)v_{x,i}(0)\rangle+ \langle v_{y,i}(t)v_{y,i}(0)\rangle\right] dt}\, \nonumber\\
\label{Dperp}
\end{eqnarray}
using standard discretization techniques \cite{HockneyEastwood}.
The length of the simulation $t_{\textrm{run}}$ was chosen to ensure the convergence of the Green-Kubo calculation, i.e. to ensure that the time integral in Eqs.~(\ref{Dparallel})-(\ref{Dperp}) reached a plateau value.

\subsection{Temperature Anisotropy Relaxation Rate}

\begin{figure*}
\includegraphics[width=5.8cm]{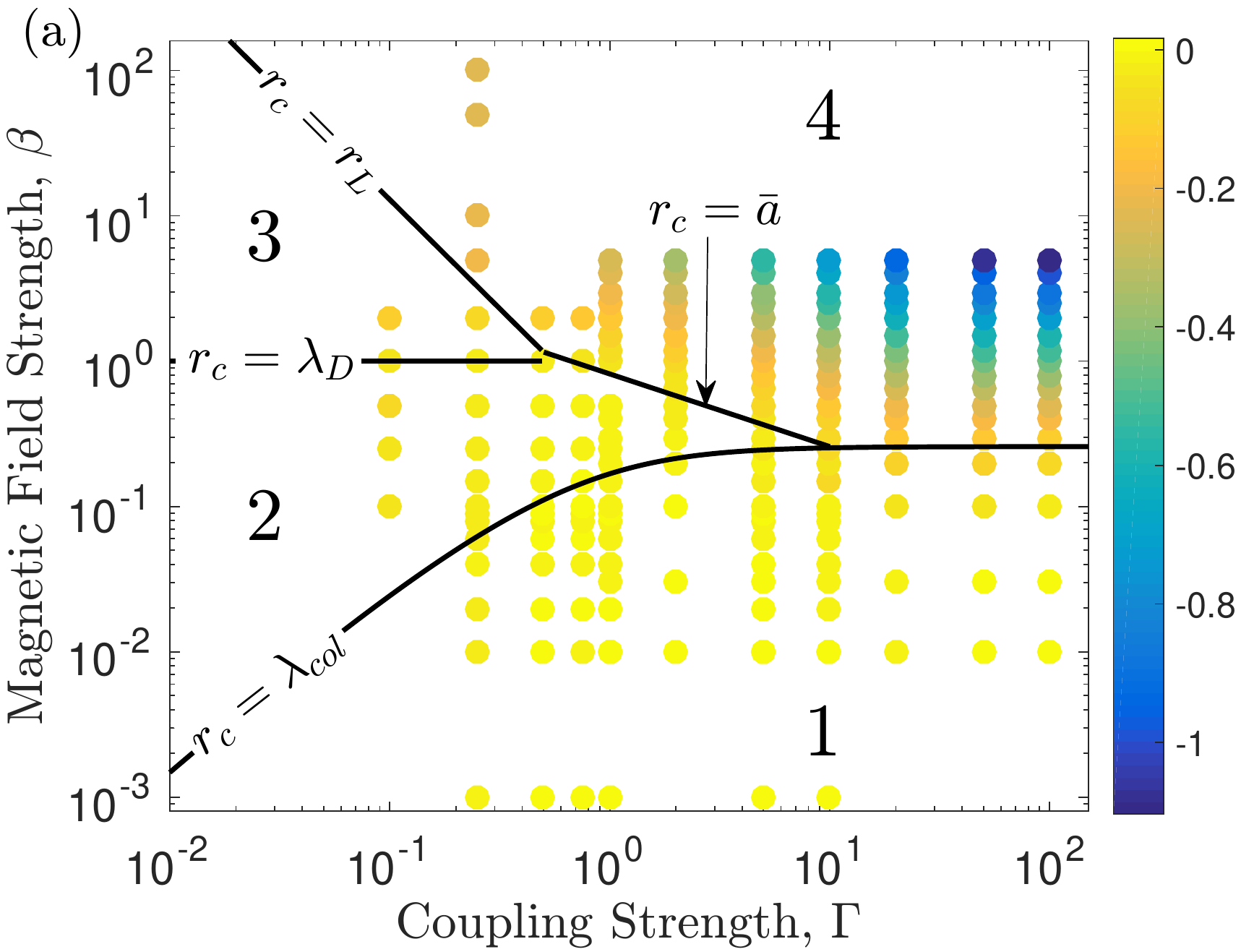}
\includegraphics[width=5.8cm]{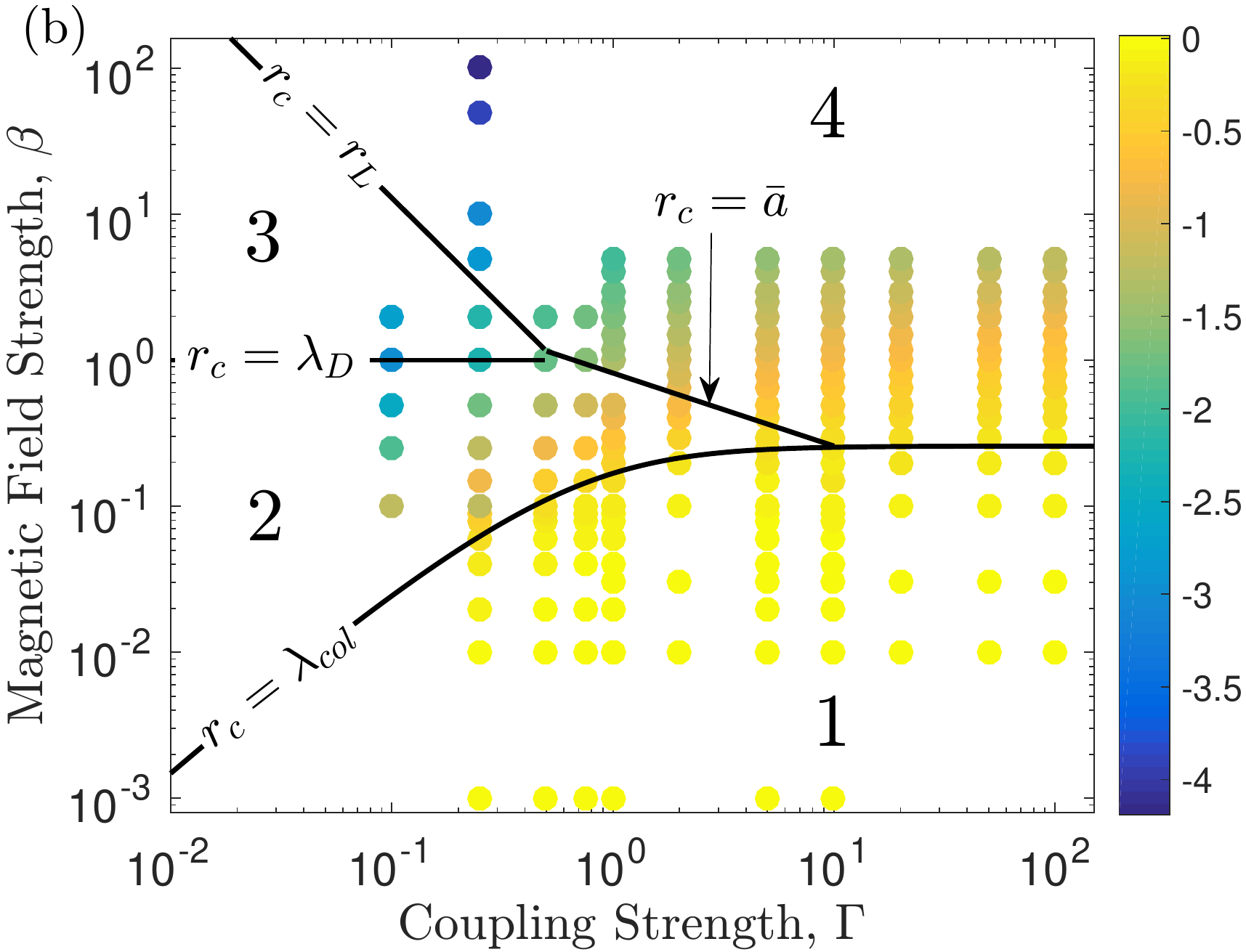}
\includegraphics[width=5.8cm]{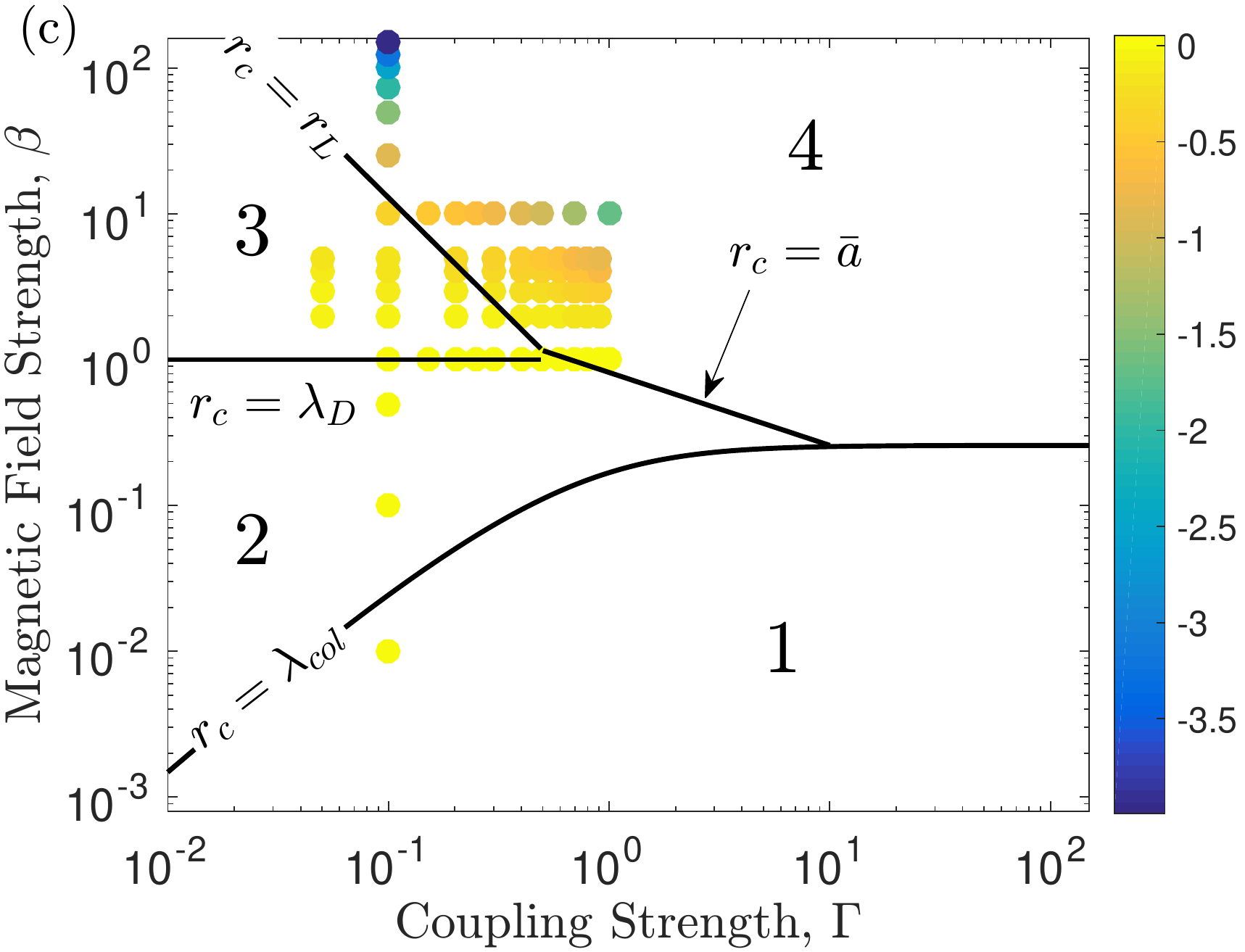}
\caption{(color online) Four transport regimes in the magnetic field strength ($\beta$) vs coupling strength ($\Gamma$) phase space: the color of circles indicates the value of (a) $\log_{10} (D_\parallel/D_o)$, (b) $\log_{10}(D_\perp/D_o)$, and (c) $\log_{10} (\nu/\nu_o)$ obtained from MD simulations. 
}
\label{fg:regimes}
\end{figure*}

At $t=0$, velocity scaling was turned off, the external magnetic field ${\bf B}=B\hat{z}$ along the $\hat{z}$-direction was turned on, and the particle velocities were rescaled to the desired initial parallel and perpendicular temperatures as follows
\begin{subequations}
\begin{eqnarray}
v_{i,x}(t=0^+)&=&v_{i,x}(0^-)\sqrt{T_\perp/T}\\
v_{i,y}(0^+)&=&v_{i,y}(0^-)\sqrt{T_\perp/T}\\
v_{i,z}(0^+)&=&v_{i,z}(0^-)\sqrt{T_\parallel/T} .
\end{eqnarray}
\end{subequations}
The system was then left to evolve freely, i.e. in the microcanonical ensemble, for a duration $t_{\textrm{run}}=300/\omega_p$.
During this period the instantaneous parallel and perpendicular temperatures, defined in terms of the particle kinetic energies
\begin{subequations}
\begin{eqnarray}
T_\parallel(t)&\equiv&\frac{2}{Nk_B}\sum_{i=1}^N{\frac{1}{2}mv_{i,z}^2(t)}\\
T_\perp(t)&\equiv&\frac{1}{Nk_B}\sum_{i=1}^N{\frac{1}{2}m\left(v_{i,x}^2(t)+v_{i,y}^2(t)\right)}\,,
\end{eqnarray}
\end{subequations}
were monitored.
In order to sample the initial statistical distribution function
\begin{equation}
f = \frac{n \exp \bigl(-\frac{mv_z^2}{2k_BT_\parallel}\bigr) \exp \bigl[ -\frac{m(v_x^2+v_y^2)}{2k_BT_\perp}\bigr]}{\pi^{3/2} \left(2k_BT_\parallel/m\right)^{1/2} \left(2k_BT_\perp/m\right)} 
\end{equation}
and to compare the MD simulations to the predictions of kinetic theories, $N_{\textrm{sim}}=25$ independent runs were performed with different initial conditions and averaged.
The averaging smooths out the fluctuations in the kinetic energies inherent to microcanonical dynamics.

To obtain the relaxation rate $\nu$, we assume that the temperatures evolve according to the rate equations,
\begin{equation}
\frac{dT_\perp}{dt} = - \frac{1}{2} \frac{d T_\parallel}{dt} = - \nu (T_\perp - T_\parallel)\,. \label{eq:nrl}
\end{equation}
The latter implies $\frac{d}{dt}\Delta T=-3\nu \Delta T$ with $\Delta T=T_\parallel-T_\perp$.
In general, as predicted by kinetic theory, $\nu$ depends on the time $t$ through its dependence upon  $T_\parallel$ and $T_\perp$.
If one assumes that the dependence is weak enough that $\nu$ is constant on a short enough time scale $\Delta t$ beyond an initial time $t_0$ , then $\Delta T(t)=\Delta T(t_0)e^{-3\nu (t-t_0)}$ on this time scale (more sophisticated treatments that take into account the temperature dependence of $\nu$ are possible (e.g. \cite{dimo:08}) but we find that they do not affect the results in any significant manner).
The MD simulations confirm that $T_\parallel-T_\perp$ indeed decays exponentially for coupling parameters $\Gamma\!<\!1$ but only following a short transient period of duration $t_0$ of order $\omega_{p}^{-1}$ (for illustration, see Fig.~4b in \cite{baal:17}). For $\Gamma\geq 1$, as discussed in the main text, the temperature evolution is never exponential and the notion of relaxation rate defined based on the rate equations (\ref{eq:nrl}) does not apply.
The initial transient period describes the dependence on initial correlations, which are discarded in the kinetic theories.
In practice, the relaxation rate $\nu$ was obtained by fitting the MD data to the analytical solution $\Delta T(t_0)e^{-3\nu (t-t_0)}$ over the time interval $[t_0,t_0+\Delta t]$ with the time $t_0$ chosen right after the early transient behavior (the data shown are obtained with $t_0=1/\omega_p$) and with $t_0+\Delta t$ chosen where the exponential behavior switches slope as a consequence of the dependence of $\nu$ on the time $t$ (Figure~4a in \cite{baal:17} illustrates this procedure when $B=0$).

\section{Results~\label{sec:results}}

\subsection{Identification of Regime Boundaries}

Figure~\ref{fg:regimes} shows that the MD results confirm the predicted regime boundary locations and quantify the smooth transitions between them. The color of the circles indicates the magnitude of $D_\parallel/D_o$ (panel a), $D_\perp/D_o$ (panel b), and $\nu/\nu_o$ (panel c) on a logarithmic scale. 
Here, $D_o$ is the self-diffusion coefficient and $\nu_o$ the temperature anisotropy relaxation rate obtained from MD simulations of the unmagentized OCP at each value of $\Gamma$. 
As predicted, when qualitative changes in transport coefficients occur, they do so at regime boundaries. 
For example, the uniformly yellow color of the circles throughout region 1 indicates that each of the transport coefficients is equal to its value in the unmagnetized OCP.
As expected, deviations occur in some of the transport coefficients, but not all, as different boundary lines are crossed. 
For example $D_\perp/D_o$ changes profoundly when the $r_c=\lambda_{\textrm{col}}$ line is crossed, whereas $D_\parallel$ changes in the transition from region 1 to 4, but not in the transition from region 1 to 2.
Each boundary can be distinguished by qualitative changes in one or more of the transport coefficients. 
The most difficult transition to distinguish on Fig.~\ref{fg:regimes} is that from regions 2 to 3, but, as seen below, it is easily identified via a change of scaling with $\beta$.

\subsection{Self-Diffusion} 

\begin{figure*}
\includegraphics[width=18cm]{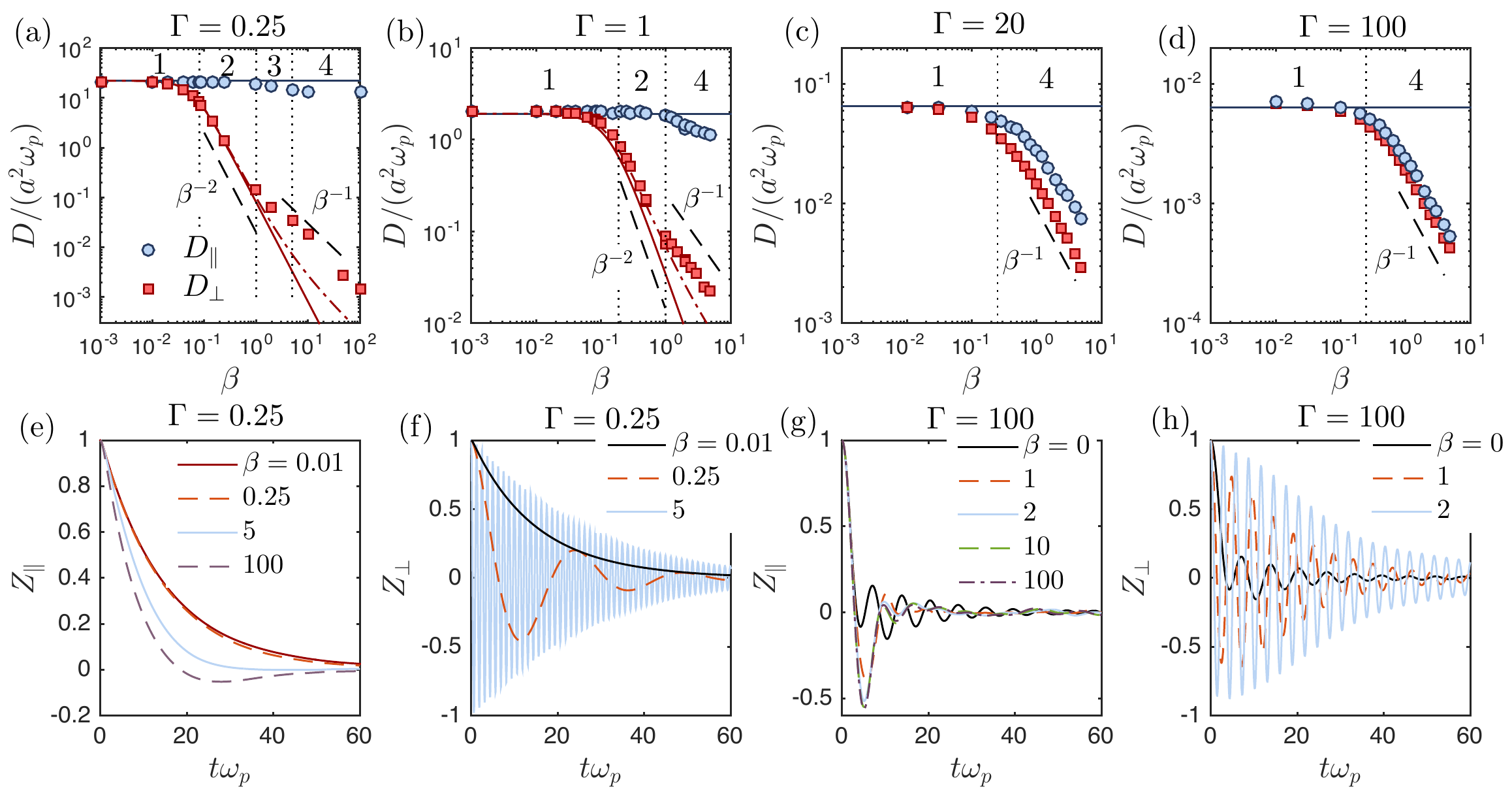}
\caption{(color online) (a)-(d) $D_\parallel$ (circles) and $D_\perp$ (squares) obtained from MD simulations. 
Solid lines in (a) and (b) show predictions of the EPT theory, and the dash-dotted lines show the predictions of the coupled mode theory of Marchetti \emph{et al} from Eq.~(\ref{eq:d_cm}). 
Solid lines in (c) and (d) show $D_o$ obtained from MD. 
The vertical dotted lines indicate the boundaries from Fig.~\ref{fg:regimes}. Panels (e)-(h) show the velocity autocorrelation functions at $\Gamma=0.25$ (e,f) and $\Gamma=100$ (g,h).}
\label{fg:diff_b}
\end{figure*}

Figure~\ref{fg:diff_b}a shows the first test of Braginskii transport theory using MD simulations.  
For the OCP, this theory predicts that to lowest order in the distribution function expansion,
\begin{eqnarray}
\frac{D_\parallel}{a^2\omega_p}&=& \frac{\sqrt{\pi/3}}{\Gamma^{5/2}} \frac{1}{\Xi}  \label{eq:d_par}
\end{eqnarray}
and
\begin{eqnarray}
D_\perp &=& D_\parallel\biggl(1 + 3\pi \frac{\beta^2}{\Gamma^3 \Xi^2} \biggr)^{-1} . \label{eq:d_perp}
\end{eqnarray}
At weak coupling, $\Xi \rightarrow \ln \Lambda$ where $\Lambda = 1/(\sqrt{3}\Gamma^{3/2})$, and this can be extended to higher coupling by computing $\Xi$ using the EPT~\cite{baal:13,baal:15}.
Figure~\ref{fg:diff_b}a shows excellent agreement between MD and the Braginskii theory in regions 1 and 2, including the independence of $D_\parallel$ on $B$ and the $B^{-2}$ scaling of $D_\perp$ in region 2 (at $\Gamma=0.25$, $\Xi \approx \ln \Lambda$).
Figure~\ref{fg:diff_b}b shows for the first time that the Braginskii theory can be extended to moderate coupling using the EPT.

The transition from region 2 to 3 shows a surprising new result. 
Standard kinetic theories addressing region 3 predict that the only change from the traditional Braginskii result is that the gyroradius sets the maximum impact parameter, modifying the Coulomb logarithm from $\ln (\lambda_D/r_L) = \ln \Lambda$ to $\ln(r_c/r_L) = \ln (\Lambda/\beta)$~\cite{sili:63,mont:74}. 
Since Braginskii theory predicts $D_\parallel \propto 1/\ln \Lambda$ and $D_\perp \propto \ln \Lambda$ in region 3, and $\beta > 1$ in this regime, a change from $\ln \Lambda$ to $\ln(\Lambda/\beta)$ implies that these standard theories predict an increase in $D_\parallel$ and a decrease in $D_\perp$ compared to the nominal Braginskii prediction (solid lines).  
It is interesting to observe that the MD data show the opposite trend. 
This calls into question the ability of these theories to address diffusion. 
It may be evidence for the recollision mechanism proposed by Dubin~\cite{dubi:14}, but we are unable to directly compare with this because it considered velocity-space diffusion rather than spatial diffusion. 

Region 3 is also observed to be consistent with Bohm scaling ($D_\perp \propto B^{-1}$)~\cite{bohm:49}.
Bohm scaling has been observed for decades in plasma experiments, notably in magnetic fusion experiments, but it is usually attributed to wave-particle scattering due to instabilities or turbulence~\cite{spit:60,wein:71}, or to collisions between charged particles and neutrals~\cite{lee:15}. 
Since our simulations were conducted at equilibrium, the plasma was stable, and the transport was solely due to Coulomb collisions. 
Bohm scaling of $D_\perp$ has been previously observed in simulations of a guiding center model of 2D systems by Taylor and McNamara~\cite{tayl:71}. 
It was also observed in 2D and 3D particle simulations by Dawson \emph{et al} \cite{daws:71,okud:72} at the highest magnetic fields simulated. 
However, the latter simulations also observed a region where $D_\perp$ was independent of magnetic field $(\propto B^0)$ over a rage of magnetic fields that would correspond to our predicted region 3. 
We did not observe such a $D_\perp \propto B^0$ region at high field in our simulations (only in region 1). 
We note also that, due to computational limitations, these early simulations were limited to small numbers of particles. 

Theories have been proposed that predict Bohm scaling at high magnetic field strength in weakly coupled plasmas. 
However, we are unaware of any that are able to provide quantitative agreement with the data in Fig.~\ref{fg:diff_b}.
For example, Marchetti \emph{et al}~\cite{marc:84} developed a mode-coupling theory, which predicts that coupling to upper hybrid modes dominates particle diffusion at sufficiently high field strength. 
The predicted perpendicular diffusion coefficient in the high field limit is $D_\perp = D_{\perp}^{(o)} + D_{\perp}^{\textrm{cm}}$, where the coupled mode contribution is 
\begin{equation}
\label{eq:d_cm}
D_\perp^{\textrm{cm}} = \frac{3\sqrt{\pi}}{2} \beta (\sqrt{3} \Gamma^{3/2})^{1/2} \biggl( \frac{\nu_c}{\omega_p} \biggr)^{1/2} D_{\perp}^{(o)} .
\end{equation} 
Here, $\nu_c$ is the Coulomb collision frequency and $D_\perp^{(o)}$ is the nominal Braginskii prediction. 
Reference~\cite{marc:84} focused on strong magnetic fields and used the high field limit of Eq.~(\ref{eq:d_perp}) for $D_\perp^{(o)}$. 
To match this with the low field limit, and to extend it to moderate coupling, we evaluate $D_\perp^{(o)}$ using Eq.~(\ref{eq:d_perp}) and the collision frequency using EPT $\nu_c/\omega_p = 0.32 \Gamma^{3/2} \Xi$. 
Figures~\ref{fg:diff_b}a and b show that, although Eq.~(\ref{eq:d_cm}) predicts Bohm scaling in region 4, the results of this theory do not quantitiatively agree with the MD results. 
We also note that this theory predicts transport regime boundaries that differ from those proposed here. 

The observation of Bohm scaling in region 3 due solely to collisional transport is an intriguing result, especially since electrons in many experiments can be found in this regime; see Fig.~\ref{fg:regimes_sketch}b. 
However, at the lowest coupling strength simulated ($\Gamma = 0.25$) region 3 is sufficiently narrow that it is difficult to draw a definitive conclusion.
Transitions between regimes are observed to be gradual, sometimes occurring over a decade in $\beta$, so it is difficult to distinguish if the observed scaling is a property of region 3, or a transition to region 4. 
We note that a recent experiment on ultracold plasma expansion in the presence of a magnetic field, where the plasma parameters are predicted to lie in region 3, could only be accurately modeled if the perpendicular diffusion coefficient was taken to have Bohm scaling (rather than the expected Braginskii scaling)~\cite{zhan:08}. 
This experiment also confirmed that no instabilities were present in the plasma that would cause significant wave-particle scattering (i.e., the transport was dominated by Coulomb collisions). 

Figures~\ref{fg:diff_b}c and d show the diffusion coefficients at strong Coulomb coupling, where regions 2 and 3 are completely absent.
The plasma transitions directly from the unmagnetized regime 1, where $D_\parallel$ and $D_\perp$ are both independent of $B$, to the extremely magnetized regime 4, where they both scale as $B^{-1}$. This regime was studied in previous 3D MD simulations by Ott and Bonitz~\cite{ott:11}. 
The data shown in Figs.~\ref{fg:regimes} and \ref{fg:diff_b}c and d are consistent with \cite{ott:11}. 
Here, we identify the boundary separating the $D_\perp \propto B^0$ and $D_\perp \propto B^{-1}$ regimes as the location where $r_c=\lambda_{col}$.
In region 4, although $D_\perp \propto B^{-1}$, the scaling of $D_\parallel$ with $B$ changes with $\Gamma$: $D_\parallel \propto B^0$ at weak coupling, but $D_\parallel \propto B^{-1}$ at strong coupling. Thus, region 4 may be properly split into 2 sub-regions separated by the $\Gamma = 1$ boundary. 
We also note that 2D MD simulations of strongly coupled plasmas have also been provided in region 4~\cite{ott:14,feng:14}.

Strong magnetization is observed to give rise to caging of particles. 
Caging is a characteristic feature of strongly coupled plasmas~\cite{donk:02}, but here it is observed to occur even at small $\Gamma$ if the magnetic field is sufficiently strong. 
Figures~\ref{fg:diff_b}e-h show the velocity autocorrelation function in the parallel ($Z_\parallel$) and perpendicular ($Z_\perp$) directions~\cite{z_footnote}. Considering $\Gamma=0.25$, in regions 1 and 2 $Z_\parallel$ monotonically decreases, as one expects for a weakly coupled plasma, and is indistinguishable from the result obtained with no magnetic field. However, as $\beta$ is increased into regions 3 and 4, nonmonontonic features arise. At $\beta = 100$, $Z_\parallel$ even takes negative values for a range of time.
In the unmagnetized OCP, such features are typically associated with the liquid-like regime (negative values occur for $\Gamma\gtrsim 50$ \cite{Daligault2006}), but here they are observed at $\Gamma=0.25$. 

The strong magnetic field effectively reduces the degrees of freedom for particle motion.
Videos of particle trajectories at these strong magnetic fields reveal that particles are constrained to move in nearly 1D, making 180$^\circ$ collisions with particles on closely neighboring field lines~\cite{sup_footnote}.
The result is that particles are essentially confined to elongated cages. 
In the direction perpendicular to $\vc{B}$ this is defined by the gyroradius, and in the direction parallel to $\vc{B}$ by the distance to a nearest neighbor particle that is on a field line close enough to Coulomb scatter. 
The resulting strong scattering by nearest neighbors and caging are physical effects similar to what is seen in an unmagnetized plasma due to strong Coulomb coupling. 
It is also noteworthy that the standard theoretical prediction that $\ln \Lambda \rightarrow \ln (\Lambda/\beta)$ suggests that the breakdown of weakly coupled theory occurs at a lower value of $\Gamma$ if the magnetic field is strong. 
The $Z_\perp$ profiles show data indistinguishable from $Z_\parallel$ at the lowest $\beta$ value, but that rapid oscillations on the timescale of $\omega_c^{-1}$ become a dominant feature for $\beta > 1$. 

Panels g and h show that at high $\Gamma$ strong magnetization further exaggerates features of the velocity autocorrelation function normally associated with Coulomb coupling, such as the large negative correlation at early times. 
Previous theories have discussed velocity caging by strong magnetization at weak coupling~\cite{Dubin1998,dubi:14}, and similar features of the velocity autocorrelation function have been observed in previous MD simulations at strong coupling~\cite{ott:11,ott:14,dzhu:16}. 
Features associated with the oscillation spectrum at strong coupling have also been observed in MD simulations~\cite{ott:12}. 

\subsection{Temperature Anisotropy Relaxation} 

\begin{figure}
\includegraphics[width=8.5cm]{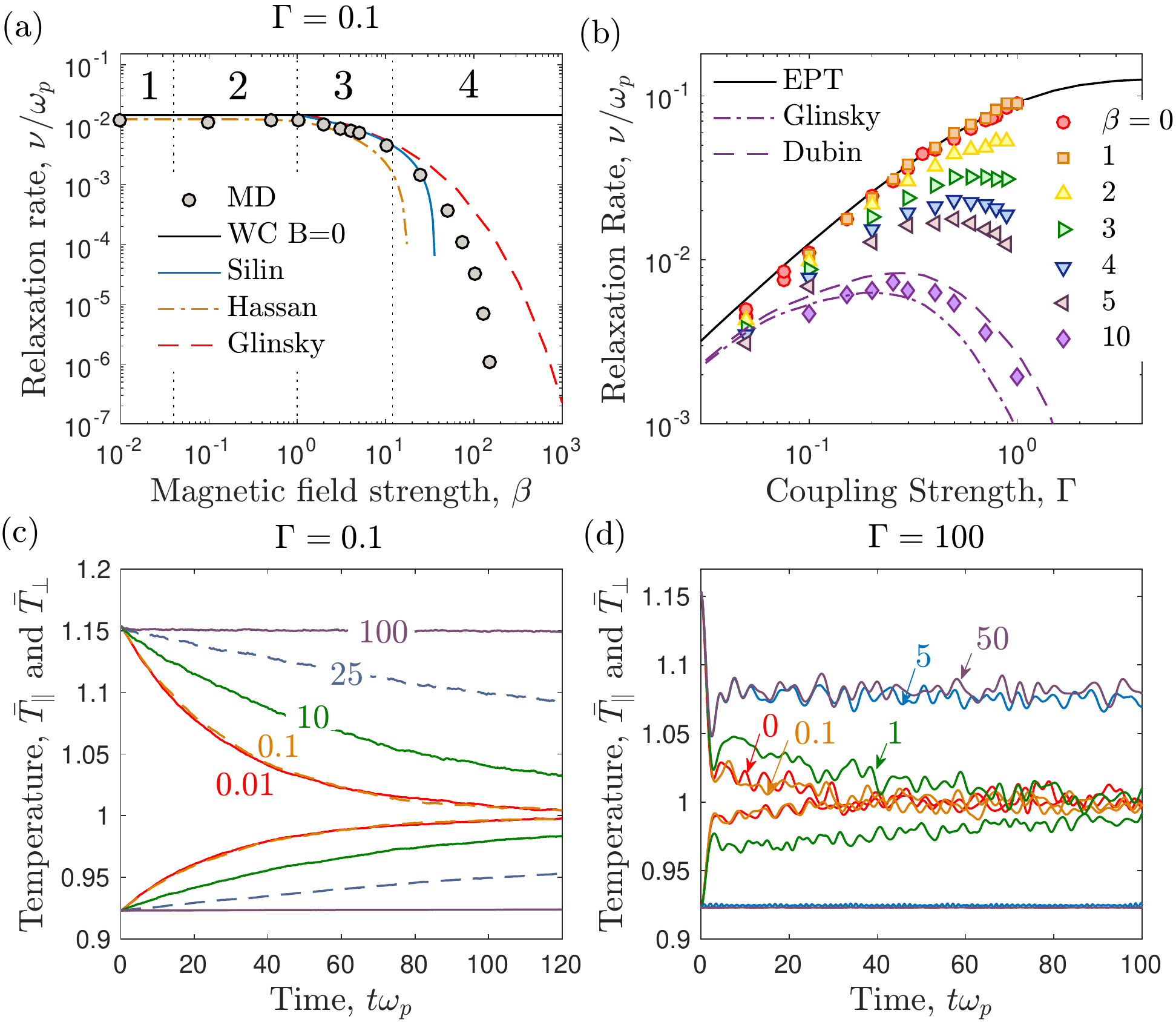}
\caption{(color online) (a) Temperature anisotropy relaxation rate vs magnetization strength at $\Gamma=0.1$ from the standard theories (lines) and MD (circles). (b) Temperature anisotropy relaxation rate vs coupling strength from MD simulations (symbols) and theory (lines). (c) and (d) Parallel and perpendicular temperature time profiles from MD for $\Gamma=0.1$ and $100$ at fixed values of $\beta$ indicated on the figure. }
\label{fg:temp}
\end{figure}

Surprisingly, the same kinetic theories that fail to describe the diffusion coefficients in region 3, accurately predict the temperature anisotropy relaxation rate.
Each of the four predicted regimes can be identified in Fig.~\ref{fg:temp}a.
The MD data are compared with four theoretical predictions.
Ichimaru et al. derived a generalized Lenard-Balescu equation for weakly coupled plasmas in uniform magnetic fields \cite{ichi:70}.
In the limit of weak magnetization and static screening, this predicts that $\nu$ is independent of $B$ in regions 1 and 2, with 
\begin{equation}
\label{eq:nu}
\nu = \alpha \sqrt{3/(4\pi)} \omega_p \Gamma^{3/2} \ln \Lambda  ,
\end{equation}
where $\alpha = (1+\frac{2}{3}A)^{3/2}[-3 + (A + 3) \arctan (\sqrt{A})/\sqrt{A}]/A^2$, and $A=T_\perp/T_\parallel - 1$; for $A\ll 1$, $\alpha \simeq 4/15$.
The prediction of this model, which is provided in standard formularies \cite{huba:16}, is shown as a black line, and agrees well with the MD results throughout regions 1 and 2.

Silin \cite{sili:63} was the first to address region 3 specifically for temperature anisotropy, predicting that the standard theory would be modified by replacing $\ln \Lambda$ by $\ln (\Lambda/\beta)$.
Hassan~\cite{Hassan1977} extended a separate Lenard-Balescu equation based model to strong magnetic fields. 
Each of these is shown to reproduce well the simulated relaxation rate in region 3. 
Here, the curve labeled ``Silin'' was computed from Eq.~(\ref{eq:nu}), but replacing $\ln \Lambda$ by $\ln (\Lambda/\beta)$. 
The curve labeled ``Hassan'' was computed using Eq.~(\ref{eq:nu_bs}), described Appendix~\ref{appendix}. 

Glinsky et al \cite{glin:92} provided a theory for weakly coupled plasmas with high magnetization, $\beta > 1$, based on O'Neil's generalized Boltzmann equation \cite{onei:83}. A prediction from this theory is shown to accurately reproduce the MD data in region 3. 
In region 4, the rate predicted by both the theory and MD decline very rapidly with $B$, but the MD data declines more rapidly. 
The curve was obtained using Eq.~(1) of \cite{glin:92} and determining $I(\bar{\kappa})$ by interpolating the data provided in tables 1 and 2. 
We note that the theory from \cite{glin:92} has been independently validated in nonneutral plasma experiments at weak coupling~\cite{beck:96}. The comparative inaccuracy with the MD simulations at $\Gamma = 0.1$ may provide further evidence for the onset of strong coupling-like behavior due to strong magnetization. 

Figure~\ref{fg:temp}b shows that magnetization strongly reduces the relaxation rate when $\beta \gtrsim 1$, and that this strong reduction onsets at lower $\Gamma$ values for higher $\beta$. 
The EPT prediction from \cite{baal:17} is found to accurately predict $\nu$ for $\beta \lesssim 1$ over this range of $\Gamma$. 
A description of how the EPT curve was obtained is provided in Appendix~\ref{appendix}. 
The conditions at which $\beta$ influences the relaxation rate are found to be consistent with the predicted boundary between regions 3 and 4 indicated in Fig.~\ref{fg:regimes}c. 
Dubin has extended the Glinsky theory to address moderate coupling, and made an analogy between the modification and the Salpeter enhancement factor of nuclear reaction rates~\cite{dubi:05}. This was later valdiated experimentally~\cite{ande:09}. 
The figure shows a comparison with MD simulations for $\beta = 10$, showing good agreement with~\cite{dubi:05}. 
The curve labeled ``Dubin'' was obtained by multiplying the ``Glinsky'' curve by the ``Salpeter enhancement factor'' $f(\Gamma)$, which was obtained as in Eq.~(3) of \cite{ande:09}, from $\ln f(\Gamma) = 1.148\Gamma - 0.00944 \Gamma \ln \Gamma - 0.000169\Gamma(\ln \Gamma)^2$; see~\cite{ichi:93}.

The temperature profiles shown in Figs.~\ref{fg:temp}c and d show non-exponential damping and oscillations indicative of correlations at strong coupling. At weak coupling the magnetic field primarily influences the rate of the monotonically merging temperatures, until there is essentially no relaxation over the timescale shown at $\beta = 100$. Much more structure is observed in the temperature profiles at strong coupling, where a rapid initial relaxation is followed by a strong arrest of the relaxation rate accompanied by oscillations (recall that each line indicates the average of 25 independent simulations). 
These effects are indicative of strong coupling, and are the reason why the temperature relaxation rate no longer follows the expected exponential behavior underlying the analysis for extracting its value. 
Recent MD simulations~\cite{ott:17} addressed strong coupling using a different approach where an initial perturbation spontaneously generated a temperature anisotropy. The relaxation rate was found to be independent of $\beta$ for $\beta \lesssim 0.3$, and exponentially decreasing with $\beta$ for larger values; see Fig.~2 of~\cite{ott:17}. This result agrees well with the predicted regime boundary in Fig.~\ref{fg:regimes}. 

\subsection{Thermal Conductivity} 

Although this work focuses on diffusion and temperature anisotropy relaxation, the phase-space of Fig.~\ref{fg:regimes_sketch} is expected to hold for other transport processes as well. Recent MD simulations of thermal conductivity~\cite{ott:15}, $\lambda$, are consistent with this prediction.
Figures~5 and 6 of \cite{ott:15} show that at strong coupling ($\Gamma = 40$) both $\lambda_\parallel$ and $\lambda_\perp$ obtain a $\beta$ dependence for $\beta \gtrsim 0.1$, consistent with crossing from regime 1 to 4. At lower coupling ($\Gamma = 5$), $\lambda_\perp$ transitions at a similar value of $\beta$, consistent with crossing from 1 to 2, while $\lambda_\parallel$ transitions at a larger value of $\beta \gtrsim 0.5$, consistent with crossing from 2 to 4. 

\section{Conclusions}

In summary, the parameter space of coupling and magnetization can be split into four regimes in which transport properties fundamentally differ. 
MD simulations of self-diffusion and temperature anisotropy relaxation agree with the predicted boundaries and quantify the smooth transitions between regimes. 
Comparison with existing theories provided validation in some regimes, such as the first test of Braginskii transport theory in the magnetized plasma regime, but also revealed disagreements with other theories and that there are regimes where no theory has yet been developed, particularly in the strong and extreme magnetization regimes. 
These results serve both as a guide to determine transport regime boundaries in plasmas spanning magnetization and coupling strength, as well as an impetus to further investigate transport properties in regimes in which accurate theory has not yet been developed. 
Figure~\ref{fg:regimes_sketch}b shows that ultracold and non-neutral plasma experiments lie in a particularly interesting regime of density and temperature because they may be able to access each of the four identified regimes at a comparatively modest magnetic field strength.  
Such experiments may provide tests of the results presented here in the future. 

\appendix \section{Evaluation of Theory Curves}
\label{appendix}

The EPT curves used to obtain the $r_c =\lambda_{\textrm{col}}$ lines in Figs.~\ref{fg:regimes_sketch} and \ref{fg:regimes}, and the self-diffusion curves in Figs.~\ref{fg:diff_b}a and b were computed using the method and formula described in \cite{baal:13,baal:15}. For completeness, these formula are provided here. The generalized Coulomb logarithm is 
\begin{equation}
\Xi^{(l,k)} = \frac{\chi}{2} \int_0^\infty d\xi\, \xi^{2k+3} e^{-\xi^2} \bar{\sigma}^{(l)} / \sigma_o \label{eq:xilk}
\end{equation}
in which
\begin{equation}
\bar{\sigma}^{(l)} = 2\pi \int_0^\infty db\, b[1 - \cos^l (\pi - 2 \Theta)] \label{eq:sig_l}
\end{equation}
is the $l^\textrm{th}$ momentum scattering cross section and $\theta = \pi - 2\Theta$ is the scattering angle, where
\begin{equation}
\Theta = b \int_{r_o}^\infty dr\, r^{-2} [1-b^2/r^2 - (e\phi(r)/k_BT)/\xi^2]^{-1/2}  . \label{eq:theta}
\end{equation}
Here, $\sigma_o \equiv \pi e^4/(2k_BT)^{2}$, $\xi^2 \equiv \frac{1}{2} u^2/v_{T}^2 = u^2/(4k_BT/m)$. Note that $\Xi \equiv \Xi^{(1,1)}$ above. The $\chi$ term is a small correction that accounts for the statistical effective size of particles using an analogy with the Enskog kinetic equation. Following~\cite{baal:15}, this is computed from
\begin{equation}
\chi \simeq 1 + 0.6250 b \rho + 0.2869 (b\rho)^2 \label{eq:chi}
\end{equation}
where $b\rho \simeq \pi n \bar{\sigma}^3/3$ and $\bar{\sigma}$ is an effective particle diameter, which is computed from the location where $g(r=\bar{\sigma}) = 0.87$. For the OCP over the range of parameters of interest here, $\chi$ ranges from 1 to 1.4; see \cite{baal:15}. 

Equations~(\ref{eq:xilk})--(\ref{eq:chi}) provide a closed set of
equations once the interaction potential, $\phi(r)$, is
specified. This was taken to be the potential of mean force $\phi(r) =
- \ln [g(r)]$, where $g(r)$ is the radial distribution function, which
was computed from the hypernetted chain (HNC) approximation
\begin{subequations}
\label{eq:hnc}
\begin{align}
g(\vc{r}) &= \exp[- v(\vc{r})/k_BT + h(\vc{r}) - c(\vc{r})] \\ 
\hat{h} (\vc{k}) &= \hat{c}(\vc{k}) [1 + n\hat{h}(\vc{k})] .
\end{align}
\end{subequations}
Here,  $v(\vc{r})/k_BT = \Gamma a/r$ is the bare Coulomb potential, $h(\vc{r}) \equiv g(\vc{r}) - 1$ and ``hats'' denote Fourier transforms in the spatial coordinate.

The EPT curve in Fig.~\ref{fg:temp}b was computed using a formula for the temperature anisotropy relaxation rate recently worked out in \cite{baal:17}:
\begin{align}
\label{eq:nu_ept}
\frac{\nu_\textrm{EPT}}{\bar{\nu}} &= \chi \frac{3 \sqrt{\pi}}{16} \frac{(1+\frac{2}{3}A)^{3/2}}{\sqrt{\alpha} A^{5/2}} \times \\ \nonumber
& \int_0^\infty d\xi\, \xi^2 e^{-\alpha \xi^2} \frac{\bar{\sigma}^{(2)}}{\sigma_o} \biggl[ \frac{2}{3} \xi^2 \alpha A \erf (\xi \sqrt{\alpha A}) - \psi (\xi^2 \alpha A) \biggr] 
\end{align}
where $\alpha \equiv T/T_\perp =  \frac{1}{3} (3+A)/(1+A)$, and
\begin{equation}
\psi(x) \equiv \erf( \sqrt{x}) - \frac{2}{\sqrt{\pi}} \sqrt{x} e^{-x}
\end{equation}
is the Maxwell integral. For the data shown in the figure, $A=-0.2$. 

The curve in Fig.~\ref{fg:temp}a labeled `Hassan'' was computed from a formula adapted from~\cite{Hassan1977} in the static screening limit. This can be written
\begin{equation}
\frac{\nu}{\nu_o} = \int \frac{dk}{k} \frac{1}{|\hat{\varepsilon} (k)|^2} \int_0^\infty d\tau\, J(\tau, s) \label{eq:nu_bs}
\end{equation}
where $\nu_o = 2ne^4 \sqrt{\pi/m}/(k_BT_\parallel)^{3/2}$, and
\begin{align}
J(\tau, s) &= \frac{2}{\sqrt{\pi}} r^{3/2} e^{-\tau^2 s^2} \times \\ \nonumber 
& \biggl\lbrace \biggl(1 - 3 \frac{r}{r-s} \biggr) \biggl[ \frac{\sqrt{\pi}}{2(r-s)^{3/2}} \frac{\erf (\sqrt{r-s} \tau)}{\tau^3}  \\ \nonumber
&- \frac{\exp [-(r-s) \tau^2]}{(r-s) \tau^2} \biggr] + 2 \frac{r}{r-s} \exp [-(r-s) \tau^2] \biggr\rbrace .
\end{align}
Here, $r = T_\parallel/T_\perp$, and  $s$ is a function of $\tau$ 
\begin{equation}
s \equiv \biggl(\frac{\sin (\tau/\kappa)}{\tau/\kappa} \biggr)^2
\end{equation}
where $\kappa \equiv 2k r_c$ and
\begin{equation}
r_{c\perp} = \frac{1}{\omega_c} \sqrt{\frac{k_BT_\perp}{m}}
\end{equation}
is the cyclotron radius using the perpendicular temperature. Here, we use the Debye-Huckel static screening 
\begin{equation}
\hat{\varepsilon} (k) = 1 + \frac{1}{k^2 \lambda_D^2}
\end{equation}
to model the dielectric function. 

\begin{acknowledgments}
This material is based upon work supported by LDRD project 20150520ER at Los Alamos National Laboratory, by the Air Force Office of Scientific Research under award number FA9550-16-1-0221 and by the U.S. Department of Energy, Office of Fusion Energy Sciences under Award Number DE-SC0016159. 
\end{acknowledgments}

\bibliography{refs.bib}

\begin{thebibliography}{99}
\bibitem{rl:footnote} The factor of $\sqrt{2}$ is included here so that $r_c=r_L$ corresponds to the regime boundary identified in \cite{glin:92}.
\bibitem{a:footnote} A $\sqrt{2}$ is included in the $r_c = a/\sqrt{2}$ boundary so that it smoothly transitions with the intersection of the $r_c=\lambda_D$ and $r_c = r_L$ boundaries.
\bibitem{z_footnote} Data for $Z_\perp$ at $\beta = 10$ and $100$ are omitted because the oscillations are so rapid on the scale shown that the lines cannot be resolved.
\end{thebibliography}

\end{document}